\documentclass[secnumarabi,superscriptaddress,amssymb,nobibnotes,aps,prc]{revtex4}

\usepackage{epsfig}
\usepackage{graphicx}
\usepackage{amsmath}
\usepackage{hyperref}

\begin{document} 

\title{Photoproduction of light vector mesons in Xe-Xe ultraperipheral collisions at the 
LHC and the nuclear density of Xe-129}

\author{V. Guzey}

\affiliation{National Research Center ``Kurchatov Institute'', Petersburg Nuclear Physics Institute (PNPI), Gatchina, 188300, Russia}
\affiliation{Department  of Physics,  University  of  Jyv\"askyl\"a, P.O. Box 35, 40014  University  of  Jyv\"askyl\"a,  Finland} 
\affiliation{Helsinki Institute of Physics, P.O.  Box  64,  00014  University  of  Helsinki,  Finland} 

\author{E. Kryshen}
\affiliation{National Research Center ``Kurchatov Institute'', Petersburg Nuclear Physics Institute (PNPI), Gatchina, 188300, Russia}

\author{M. Zhalov}
\affiliation{National Research Center ``Kurchatov Institute'', Petersburg Nuclear Physics Institute (PNPI), Gatchina, 188300, Russia}

\begin{abstract} 

We make predictions for cross sections of $\rho$ and $\phi$ vector meson photoproduction in ultraperipheral Xe-Xe collisions at $\sqrt{s_{NN}}=5.44$ TeV. Analyzing the momentum transfer distribution of $\rho$ mesons in this process,
 we explore the feasibility of extracting the nuclear density of $^{129}$Xe, 
which is needed in searches for dark matter
with Xenon-based detectors.

\end{abstract}

\maketitle 

\section{Introduction}
\label{sec:Intro}

Collisions of ultrarelativistic ions at large impact parameters -- the 
so-called ultraperipheral collisions (UPCs) -- provide opportunities 
to explore photon--photon, photon--proton, and photon--nucleus
interactions at previously unattainable high energies~\cite{Baltz:2007kq}.
In particular, a test run of collisions of Xenon ions was performed 
at the Large Hadron Collider (LHC) in Fall 2017. The experiments have collected several ${\mu}$b$^{-1}$ of 
statistics, which is sufficient to study photoproduction of light $\rho$ and $\phi$ vector mesons.
 Extending the formalism~\cite{Frankfurt:2015cwa}, which reasonably
describes coherent $\rho$ meson photoproduction in Au-Au UPCs~\cite{Adler:2002sc} 
at the Relativistic Heavy Ion Collider (RHIC) and Pb-Pb UPCs~\cite{Adam:2015gsa} 
at the LHC, we consider coherent and 
incoherent $\rho$ and $\phi$ meson photoproduction in Xe-Xe UPCs in 
the LHC kinematics at $\sqrt{s_{NN}}=5.44$ TeV and predict the corresponding 
UPC cross sections as functions of the vector meson rapidity $y$ and 
the momentum transfer $t$. 
These predictions combined with 
the earlier results for Pb-Pb UPCs provide the nuclear 
mass number $A$ dependence of our approach to nuclear shadowing in light 
vector meson photoproduction on nuclei and can be compared to the future LHC data.

In the past, 
based on the vector meson dominance (VMD) model, 
photoproduction of light vector mesons on nuclei was used 
to determine the meson--nucleon cross section
and to constrain the nuclear matter density distribution of the target~\cite{Bauer:1977iq}. 
To our knowledge, 
only the root-mean-square (rms) charge radii of Xe isotopes 
have been extracted from isotope shift
measurements~\cite{Libert:2007mrz} and the charge density distribution of
$^{132}$Xe was recently determined~\cite{Tsukada:2017llu} 
from electron-xenon elastic scattering at SCRIT facility. At the same time
 the nuclear matter distribution, the
effective nuclear radius, and the structure factors of Xe isotopes 
are of key importance 
for  Dark Matter experiments searching for 
weakly interacting massive particles (WIMP) with Xenon-based 
detectors (for details, see, e.g.~\cite{Uchida:2014cnn,Vietze:2014vsa,Fieguth:2018vob}).
In this note we demonstrate that the measurement of $\rho$ photoproduction 
in Xe-Xe UPC at the LHC
can be used to gain information on the nuclear matter distribution in $^{129}$Xe.

\section{Coherent and incoherent cross sections of $\rho$ and $\phi$ photoproduction in nucleus--nucleus UPCs}
\label{sec:main}

The cross section of coherent and incoherent (the target nucleus breaks up) 
cross section of vector meson $V$ ($V=\rho, \phi$) 
photoproduction in symmetric nucleus--nucleus UPCs reads~\cite{Baltz:2007kq}:
\begin{equation}
\frac{d\sigma_{AA \to V AA^{\prime}}(y)}{dy}=N_{\gamma/A}(y) \sigma_{\gamma A \to V A^{\prime}}(y)+
N_{\gamma/A}(-y) \sigma_{\gamma A \to V A^{\prime}}(-y) \,,
\label{eq:cs1}
\end{equation}
where $N_{\gamma/A}$ is the photon flux; $y$ is the rapidity of 
the produced vector meson $V$; $\sigma_{\gamma A \to V A^{\prime}}(y)$ is 
the photoproduction cross section.
The target nucleus label $A^{\prime}$ stands for both coherent $A^{\prime}=A$ 
and incoherent $A^{\prime} \neq A$ cases.
The presence of two terms with the opposite rapidities in Eq.~(\ref{eq:cs1}) 
reflects the fact that each colliding ion can serve
as a source of photons and as a target.

The photon flux $N_{\gamma/A}(y)$ produced by an ultrarelativistic ion in 
nucleus--nucleus UPCs in Eq.~(\ref{eq:cs1})
can be very well approximated by the photon flux due to a point-like charge $Z$:
\begin{equation}
N_{\gamma/A}(y)=\frac{2 Z^2 \alpha_{\rm e.m.}}{\pi} \left[\zeta K_0(\zeta)K_1(\zeta)-\frac{\zeta^2}{2}\left(K_1^2(\zeta)-K_0^2(\zeta)\right)\right] \,,
\label{eq:flux}
\end{equation}
where $\alpha_{\rm e.m.}$ is the fine-structure constant; $K_{0,1}$ 
are Bessel functions of the second kind;
$\zeta=\omega b_{\rm min}/\gamma_L$; $\omega=(M_V/2) e^y$ is the 
photon energy for given $y$, where $M_V$ is the vector meson mass;
$\gamma_L$ is the nucleus Lorentz factor 
in the laboratory frame; 
$b_{\rm min}$ is the minimal transverse distance between 
the centers of the colliding nuclei specifying the ultraperipheral collision. 
Its value $b_{\rm min}\approx 2R_{A}$ ($R_{A}$ is the radius of the nucleus) 
is found by requiring 
that Eq.~(\ref{eq:flux}) reproduces the photon flux, which 
is calculated as convolution over impact parameters of the flux of equivalent 
photons produced by the charge distribution of the radiating nucleus
with the probability to not have the strong inelastic
interactions in a given nucleus--nucleus collision.

In high-energy UPCs
 of heavy ions with the large charge $Z$,
the photoproduction process can be accompanied by additional 
photon exchanges between colliding ions because the 
parameter $\alpha_{\rm e.m.}^2 Z^2$ is not small.
These additional photon exchanges 
may lead to excitations of one or both 
colliding nuclei~\cite{Vidovic:1993cf,Pshenichnov:2001qd}, which typically decay by emission of 
one or more neutrons moving along the direction of ion beams and detected by zero-degree
calorimeters (ZDCs).
The low-energy electromagnetic excitation of nuclei and the 
high-energy vector meson photoproduction in UPC can be considered as
 independent processes 
because of the large difference in time scales. Hence,
one can account for the additional photon exchanges by modifying
the photon flux and, thus, selecting photoproduction of vector mesons
in nucleus--nucleus UPCs in different channels $i$, which are specified by
emission of various number of neutrons
 $i=(0n0n,1n1n,0nXn,XnXn, \dots)$~\cite{Baltz:2002pp}.
 %---
 In particular, the photon flux for channel $i$ reads:
 \begin{equation}
 N_{\gamma/A}^{i}(y)=\int_{2 R_A}^{\infty} d^2 \vec{b}\, N_{\gamma/A}(y,\vec{b}) P_i(\vec{b})
 \label{eq:flux_i}
 \end{equation}
 where $N_{\gamma/A}(y,\vec{b})$ is the photon flux at the transverse distance $b$ (impact parameter) from the center of the nucleus,
 which produces it; $P_i(\vec{b})$ is the probability to emit a given number of neutrons corresponding to channel $i$.
This approach describes very well the ALICE data on electromagnetic dissociation in Pb-Pb UPCs~\cite{ALICE:2012aa} and is implemented in the Starlight Monte Carlo generator~\cite{Klein:2016yzr}, which is commonly used for calculations and
simulations of various UPC processes.
Note that an alternative approach to electromagnetic excitation of nuclei with neutron emission in UPCs, which is based on 
the Hauser--Feshbach formalism and which  provides a good description of the RHIC and 
LHC data on electromagnetic excitations in UPCs,
was developed in~\cite{Klusek-Gawenda:2013ema}. 
%---

The coherent  $\gamma A \to V A$ cross section $\sigma_{\gamma A \to \rho A}$ 
in Eq.~(\ref{eq:cs1})
 can be calculated using the combination of 
the Gribov--Glauber model for nuclear shadowing and a model for 
hadronic fluctuations for the $\gamma N \to VN$ 
cross section~\cite{Frankfurt:2015cwa,Guzey:2013jaa}. This approach provides a 
good description of the data on coherent 
$\rho$ photoproduction on heavy nuclei in UPCs at RHIC and the LHC (Run 1). 
It is based on the observation that at high energies, the real photon interacts with hadronic targets by means of its long-lived hadronic 
components (fluctuations). Each fluctuation is characterized by 
the cross section $\sigma$ and 
interacts independently with nucleons of a nuclear target;
 the probability distribution of these
fluctuations $P(\sigma)$ is constrained using the experimental 
data on the elastic $\gamma p \to V p$ and the diffraction 
dissociation $\gamma p \to X p$ cross sections, see details 
in Ref.~\cite{Frankfurt:2015cwa,Guzey:2016piu}.
Thus, the $\gamma A \to V A$ cross section in the 
large $W_{\gamma N}$-limit ($W_{\gamma N}$ is the invariant 
photon--nucleus energy per nucleon) is given by the following expression:
\begin{equation}
\sigma_{\gamma A \to V A}^{\rm mVMD-GGM}(W_{\gamma N})=
\left(\frac{e}{f_V}\right)^2 \int d^2 \vec{b} 
\left|\int d\sigma P(\sigma) 
\left(1-e^{-\frac{\sigma}{2}\tilde{T}_A(\vec{b})}\right) \right|^2 \,,
\label{eq:cs_A}
\end{equation}
where $f_V$ is the $\gamma-V$ coupling constant ($f_{\rho}^2/4 \pi=2.01$ 
for $\rho$ and $f_{\phi}^2/4 \pi=13.7$ for $\phi$); 
$\tilde{T}_A(\vec{b})=
\int^{\infty}_{-\infty} dz \rho_A(\vec{b},z)-(l_c \sigma)/2 
\int^{\infty}_{-\infty} dz \rho_A^2(\vec{b},z) $ is the nuclear optical density, 
which also takes into
account short-range nucleon--nucleon ($NN$) correlations in the nuclear wave function,
where $\rho_A(b,z)$ is the nuclear density and $l_c=-0.74$ fm is 
the $NN$ correlation length. 
For lower values of $W_{\gamma N} \leq {\cal O}(\sqrt{2 R_A m_N} M_V) \approx 5$ GeV,
the expression in Eq.~(\ref{eq:cs_A}) should be corrected by including 
the effects of the non-zero longitudinal momentum
transfer in the $\gamma N \to V N$ amplitude (the effect of nuclear coherence). 
It suppresses the 
$d \sigma_{AA \to V AA}/dy$ UPC cross section~(\ref{eq:cs1}) 
at forward and backward rapidities but does not affect
it near $y \approx 0$.

In the case of incoherent nuclear scattering, 
the $\gamma A \to V A^{\prime}$ quasi-elastic cross section 
$\sigma_{\gamma A \to \rho A^{\prime}}$ can be calculated 
using completeness of final nuclear states $A^{\prime}$,
see, e.g.~\cite{Bauer:1977iq}. Applying the photon fluctuations 
to the nuclear scattering amplitudes, 
one obtains:
\begin{equation}
\sigma_{\gamma A \to V A^{\prime}}^{\rm mVMD-GGM}(W_{\gamma N})= 
\sigma_{\gamma N \to  V N}(W_{\gamma N}) \int d^2 \vec{b}\, 
\tilde{T}_A(\vec{b})\left|\int d\sigma P(\sigma) 
\frac{\sigma}{\langle \sigma \rangle} 
e^{-\frac{\sigma}{2}\tilde{T}_A(\vec{b})}\right|^2 \,,
\label{eq:cs_incoh}
\end{equation}
where $\langle \sigma \rangle=\int d\sigma P(\sigma) \sigma$. 
Equation~(\ref{eq:cs_incoh}) has a clear physical interpretation:
quasi-elastic photoproduction of $\rho$ mesons on a nuclear 
target corresponds to elastic $\rho$ production on 
any from
all $A$ target nucleons
with the condition that interactions with remaining nucleons 
do not lead to inelastic production.
The probability to not 
have inelastic processes 
describes the effect of nuclear shadowing for individual 
fluctuations and depends on the distribution
$P(\sigma)$. In the absence of fluctuations, it reduces to the familiar 
Glauber model expression.

In collider kinematics of 
ion UPCs at ALICE,
 it is problematic to separate
the quasielastic incoherent process $\gamma A \to V A^{\prime}$ and
photoproduction of vector mesons $\gamma A \to V A^{\prime}Y$ with nucleon 
dissociation $\gamma N \to V Y$ into not too large masses $M_{Y}< 10$ GeV. 
All fragments $Y$ are going in the very forward direction along the beams.
We estimate the contribution of this process within the Gribov--Glauber model. 
In particular, 
an examination of corresponding multiple scattering graphs shows that 
the $\gamma N \to V Y$ cross
section $\sigma_{\gamma N \to V Y}$ factorizes out and the 
remaining nuclear shadowing suppression is the same as in the case of 
Eq.~(\ref{eq:cs_incoh}). 
It can also be shown formally by generalizing the derivation of Eq.~(\ref{eq:cs_incoh}) 
to include color fluctuations~\cite{Frankfurt:2008vi} in 
target nucleons (nucleon shape fluctuations~\cite{Mantysaari:2016ykx}) 
and keeping the leading power of the variance of these fluctuations.
To this accuracy, the form of these fluctuations is not important; 
the variance is expressed in terms of $d\sigma_{\gamma_{N \to V Y}}(t=0)/dt$. 
Therefore, the incoherent cross section of light vector 
meson $V$ photoproduction on nuclei
with target
nucleon dissociation 
is given by the following expression:
\begin{equation}
\sigma_{\gamma A \to VA^{\prime} Y}^{\rm mVMD-GGM}(W_{\gamma N})= 
\sigma_{\gamma N \to  V Y}(W_{\gamma N})
\int d^2 \vec{b}\, T_A(\vec{b})\left|\int d\sigma P(\sigma) \frac{\sigma}{\langle \sigma \rangle} e^{-\frac{\sigma}{2}\tilde{T}_A(\vec{b})}\right|^2 \,.
\label{eq:cs_incoh2}
\end{equation}

Calculations of the 
cross section of $\rho$ photoproduction with dissociation of the proton target
in the non-perturbative domain of small values of $|t|$
are strongly model-dependent. Instead, for the kinematical region of ALICE 
measurements, we 
use the HERA results~\cite{Breitweg:1999jy,Weber:2006di}
and obtain 
for the ratio of the forward target-dissociative and elastic 
$\rho$ 
photoproduction cross sections on the proton:
\begin{equation}
\frac {d\sigma_{\gamma p \to \rho Y}(t\approx 0)/dt} 
{d\sigma_{\gamma p \to \rho p}(t\approx 0)/dt} \approx 0.1 - 0.12 \,.
\end{equation}
Combining this value with the ratio of 
corresponding slope parameters of 
the $t$ dependence $B_{\rm diss}/B_{\rm el} \approx 0.25$
(it is assumed that the $t$ dependence is exponential),
we obtain the following relation:
\begin{equation}
\sigma_{\gamma p \to  V Y}(W_{\gamma p})\approx
0.5\,\sigma_{\gamma p \to  V p}(W_{\gamma p}) \,.
\label{eq:diselrat}
\end{equation}
%---
Note that this relation also agrees very well with the ratio of the elastic and proton-dissociation cross sections of $\rho$ electroproduction
in a wide range of $Q^2$ and at $W_{\gamma p}=75$ GeV measured by the H1 collaboration at HERA~\cite{Aaron:2009xp}, 
which shows that Eq.~(\ref{eq:diselrat}) is approximately $Q^2$-independent.
%---

It should be emphasized that the $t$-integrated cross section of 
vector meson photoproduction with target
nucleon dissociation is comparable with the incoherent quasielastic cross section
due to the contribution of large $|t|$ and a smaller slope parameter. The contribution of this
process in the case of a heavy nuclear target is not essential for small $|t|$, 
in particular, in the region $|t|< 0.02$ GeV$^2$, where coherent photoproduction dominates
  the momentum transfer distribution. 

Before discussing results of our calculations presented in Figs.~\ref{fig:rhoxe5440}-\ref{fig:radxe0n},
 we note that the detailed
description of our approach,
the modified VMD-GGM, is given in \cite{Frankfurt:2015cwa}. Here we only 
emphasized the feature specific for the current calculations.
The main part of our results is obtained with the nuclear matter density distribution 
of $^{129}$Xe calculated within
the standard spherical Hartree--Fock--SkyrmeIII model with accounting for the 
BCS pairing (no special fit specific to the Xe nucleus was done). 
This model gives the values of rms mass radius of
$^{129}$Xe equal to 4.818 fm and the charge rms radius equal to 4.77 fm. 
The latter can be compared to the experimental 
value  $4.7831 \pm 0.0043$~\cite{Libert:2007mrz} obtained 
from the isotope shift measurements.
The effective radius $R_A =5.8$ fm, which is 
determined as the distance from
the center to the point, where the mass density decreases by a factor of two
 compared to its maximal value, is somewhat larger than the value $R_A =5.45$ fm, which is
given by the commonly
 used parametrization 
$R_{A}=1.112\, A^{1/3}-0.86\, A^{-1/3}$ fm.

\section{Results}
\label{sec:results}

\begin{figure}[ht!]
\begin{center}
\epsfig{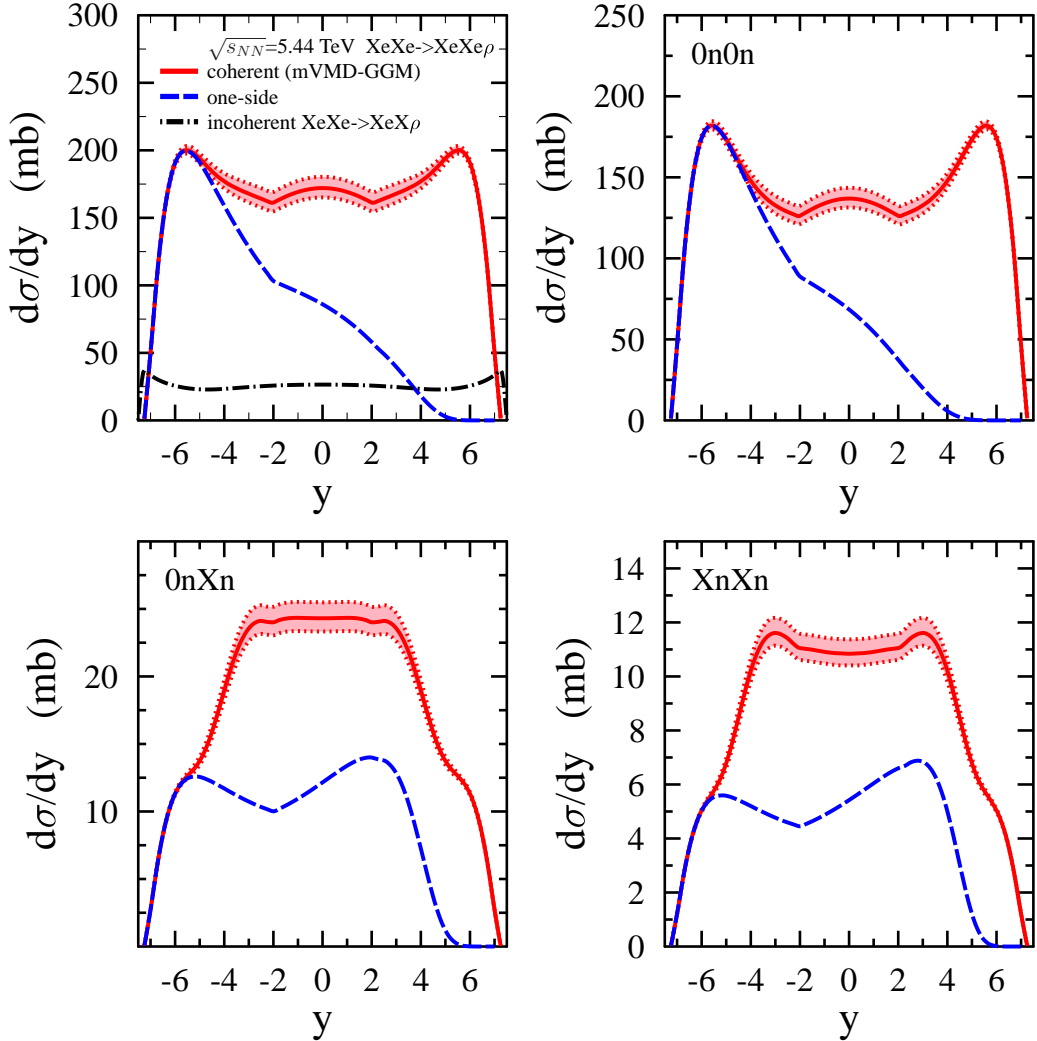}
\caption{Rapidity distributions for $\rho$ photoproduction in ultraperipheral
Xe-Xe collisions at $\sqrt {s_{NN}}=5.44$ TeV at the LHC. 
See explanations in text.}
\label{fig:rhoxe5440}
\end{center}
\end{figure}

\begin{figure}[h]
\begin{center}
\epsfig{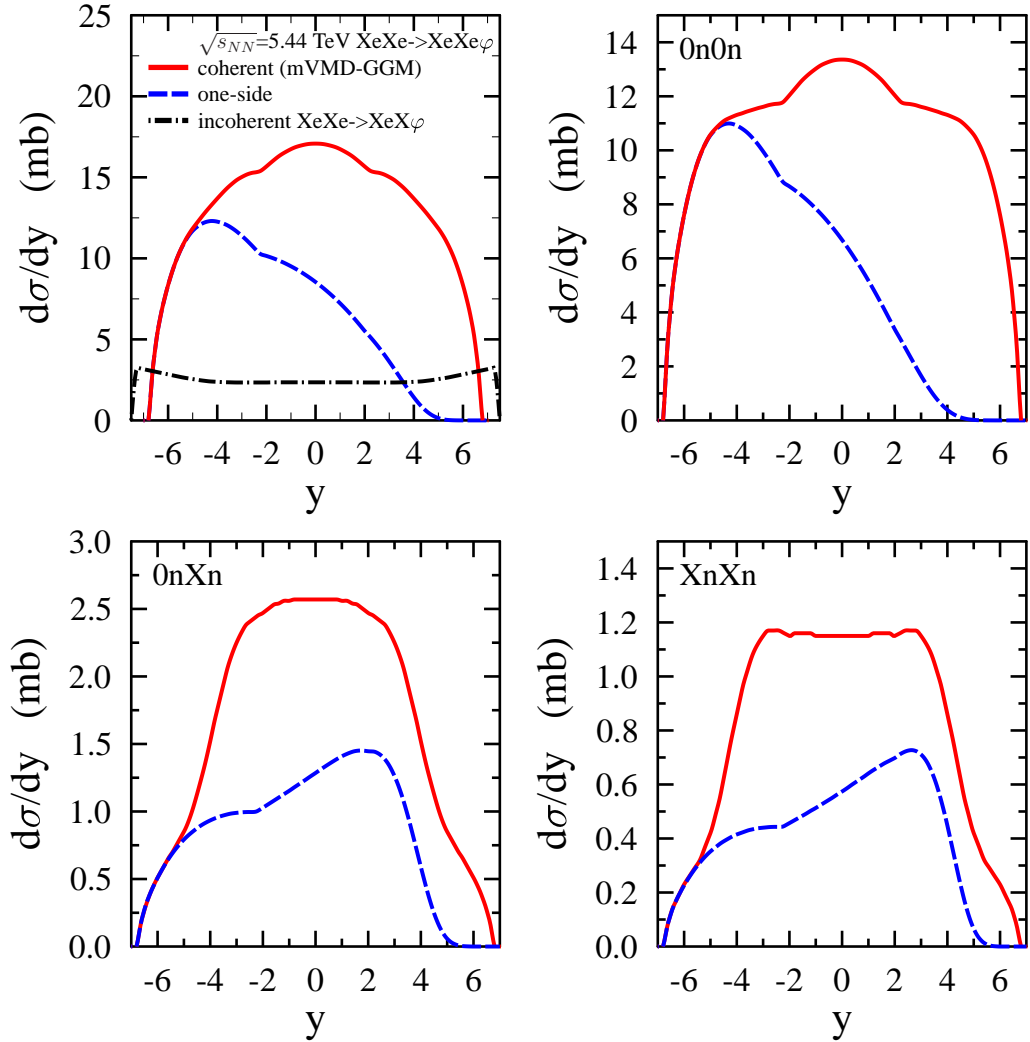}
\caption{Rapidity distributions for $\phi$ photoproduction in ultraperipheral
Xe-Xe collisions at $\sqrt {s_{NN}}=5.44$ TeV at the LHC. See explanations in text.}
\label{fig:phixexerun2}
\end{center}
\end{figure}

\begin{figure}[h]
\begin{center}
\epsfig{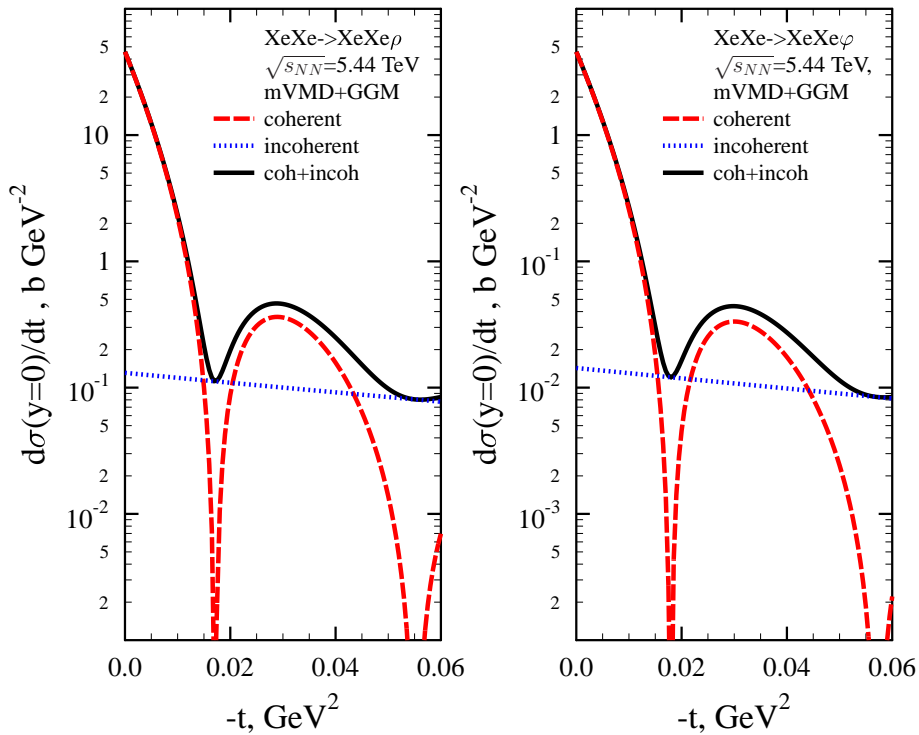}
\caption{The $t$-dependence of cross section of light vector meson 
($\rho$ on the left and $\phi$ on the right) photoproduction
in ultraperipheral Xe-Xe collisions at the LHC at $\sqrt {s_{NN}}=5.44$ TeV.}
\label{fig:xedsdt}
\end{center}
\end{figure}

\begin{figure}[h]
\begin{center}
\epsfig{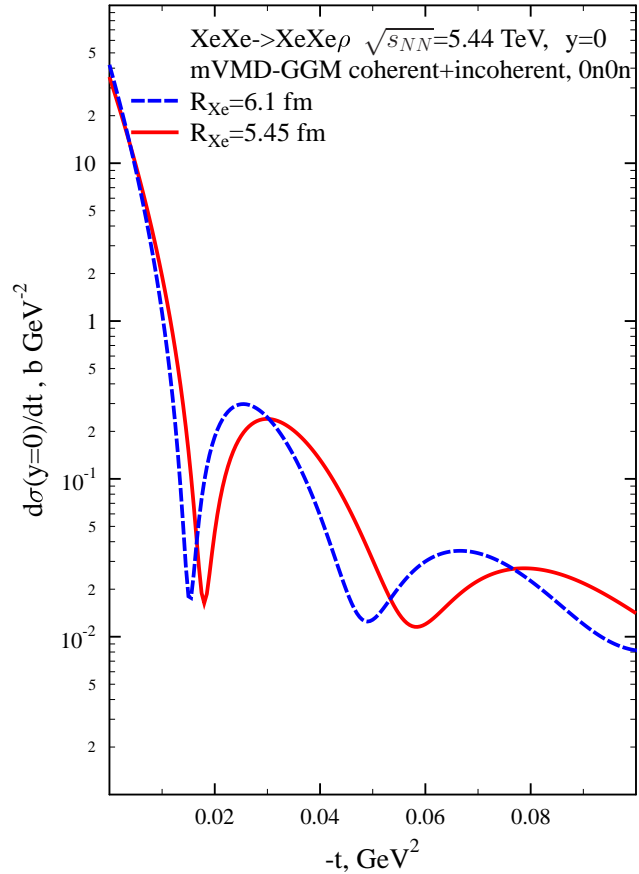}
\caption{The momentum transfer distribution of $\rho$ photoproduction
in Xe-Xe UPCs
calculated using the
mVMD-GGM approach with the two-parameter Fermi distribution
of the Xenon nuclear density and two values of $^{129}$Xe radius.}
\label{fig:radxe0n}
\end{center}
\end{figure}

The calculated rapidity distributions for $\rho$ and $\phi$ mesons 
produced in ultraperipheral Xe-Xe collisions at $\sqrt {s_{NN}}=5.44$ TeV
are presented in Figs.~\ref{fig:rhoxe5440} and \ref{fig:phixexerun2}, respectively. 
In four panels 
we show the total rapidity distribution and the rapidity distributions in different
channels: ``0n0n" 
is the case 
without additional photon exchanges, whose experimental signature is  
no neutrons detected in zero-degree calorimeters (ZDCs), the ``0nXn" 
channel corresponds
to electromagnetic excitation with subsequent neutron decay of only one of 
colliding ions, and ``XnXn" 
is the case of mutual excitation of both ions. Red solid
lines with strips show the
coherent cross section with the uncertainty of our model in accounting for
Gribov inelastic shadowing.  Blue dashed lines show the contribution 
to the rapidity distribution from one target nucleus
corresponding to the first term in Eq.~(\ref{eq:cs1}).
Finally, the black dot-dashed line presents the summed contribution of quasielastic and
nucleon target-dissociative photoproduction calculated as described above.
%---
Specifically, the black dot-dashed line is obtained by adding the contributions of Eqs.~(\ref{eq:cs_incoh})
and (\ref{eq:cs_incoh2}), where the proton-dissociation cross section $\sigma_{\gamma p \to  V Y}(W_{\gamma p})$ is calculated
using Eq.~(\ref{eq:diselrat}). 
As one can see from Eq.~(\ref{eq:diselrat}), the inclusion of the proton dissociation contribution increases the $t$-integrated
nuclear incoherent cross section by approximately 50\%.
%---
 All cross sections
in  Figs.~\ref{fig:rhoxe5440} and \ref{fig:phixexerun2}
have been integrated over the momentum transfer.

 The shape of the rapidity
distributions for coherent photoproduction reflects an interplay of several 
phenomena. Bumps at forward and backward rapidities in upper panels 
of Fig.~\ref{fig:rhoxe5440} are due to an enhanced 
contribution of low-energy photoproduction related to the secondary Reggeon exchange
in the $\rho$-N interaction; the inelastic Gribov shadowing at low energies is
still small. Since only the Pomeron exchange contributes to the $\phi -N$ interaction, 
such bumps are absent in $\phi$ photoproduction (Fig.~\ref{fig:phixexerun2}).
The one-side contribution demonstrates how an
interplay of the energy dependence of the elementary cross section,
suppression due to nuclear shadowing, and drop of the flux of high-energy equivalent photons
determine the distribution in the central and forward rapidity regions.
A comparison of the distributions in different channels shows that additional photon 
exchanges resulting in neutron decays of excited nuclei enhance the role of smaller
impact parameters of the collision~\cite{Baltz:2002pp} and, correspondingly,
enhance the high-energy contribution to the UPC cross section.

The calculated transverse momentum distributions at the rapidity $y=0$ are
presented in Fig.~\ref{fig:xedsdt}. The coherent cross section is shown 
by the red dashed line, the incoherent one -- by the blue dotted line, and the summed 
cross section is given by the solid black curve.
Here we neglected the contribution of the photoproduction process
with nucleon dissociation, whose contribution is at the level of a few percent 
in the region of small $|t|<0.1$ GeV$^2$. 
We also neglected  washing out
of the diffractive dip in the coherent cross section due to a small,
but non-vanishing transverse momentum of 
quasireal photons~\cite{Klein:1999gv} and the real part of the meson--nucleon amplitude.

It is well known that the position of diffractive dips is very sensitive to the radius 
of the target nucleus. Hence, to reveal them more clearly one needs to suppress the
incoherent contribution. 
It can be achieved by selecting the ``0n0n"
channel of photoproduction, where one requires no
forward neutron emission.
 It was shown in~\cite{Strikman:2005ze} that incoherent events 
of high energy photoproduction of vector mesons for $|t|>0.03$ GeV$^2$ are predominantly
accompanied by neutrons from the decay of the excited residual nucleus. 
In Fig.~\ref{fig:radxe0n} we show the momentum transfer distributions for the ``0n0n"
channel in photoproduction of $\rho$ meson in Xe-Xe UPCs at the rapidity $y=0$.
To reveal the possibility of study the influence of the nuclear radius,
we performed these calculations with the
two-parameter Fermi distribution of the nuclear density 
$\rho (r)=\rho_0 [1+\exp((r-R_A )/a)]^{-1}$
 with the parameters
$a=0.54$ and $R_A =6.1$ fm and with $R_A =5.45$ fm.  
It is seen that in measurements with high statistics and momentum resolution, 
by fitting the shape of the momentum transfer distribution in the region up to
$|t|\approx 0.1$ GeV$^2$, one can
determine the nuclear radius with rather high accuracy.
%---
Note that the theoretical uncertainty of our calculation of inelastic nuclear shadowing, which is shown by shaded bands in  
Fig.~\ref{fig:rhoxe5440}, affects primarily the cross section magnitude and not the shape of the $t$ dependence. 
Hence, this theoretical uncertainty does not affect positions of the minima in Figs.~\ref{fig:xedsdt} and \ref{fig:radxe0n}.
%--- 

%---
There are also alternative approaches to calculation of light vector meson photoproduction in UPCs, notably, 
the one based on the color dipole framework and the phenomenological data-driven approach, which is used in 
Starlight Monte-Carlo generator.
Briefly, 
photoproduction of light vector mesons in heavy ion Pb-Pb UPCs at the LHC has been
extensively studied in the framework of the color dipole model including saturation 
effects~\cite{Ivanov:2007ms,Goncalves:2011vf,Santos:2014vwa,Klusek-Gawenda:2016oqd}. Since the cross section of the
discussed process is very sensitive to the non-perturbative contribution of large dipoles, the dipole model predictions
strongly depend on the choice of the dipole cross section in this region and the final light vector meson wave function.
In addition, due to the sub-leading Reggeon contribution to the $\gamma p \to \rho p$ cross section, 
we predict a two-bump shape of the rapidity distribution $d\sigma_{A A \to \rho AA}/dy$, while the dipole models naturally
lead to the distribution, which is bell-shaped.

Another framework to describe photoproduction of vector mesons in ion UPCs is based on the Starlight Monte-Carlo 
generator~\cite{Klein:2016yzr}. It combines phenomenological parameterizations of the $\gamma p \to V p$ cross sections on the 
proton with the optical theorem and the classical expression for the interaction of vector mesons with nuclei
and rather successfully describes the available data on $\rho$ photoproduction
in Au-Au UPCs at RHIC and Pb-Pb UPCs at the LHC~\cite{Adamczyk:2017vfu,Klein:2017vua}.
In the context of present analysis, it is important to note that while the $t$ dependence of $d \sigma_{\gamma A \to V A}/dt$ 
is given by the nuclear form factor squared $F_A^2(t)$ in Starlight (which is true in the limit of small nuclear shadowing),
it is shifted towards smaller $|t|$ in our case. An indication of this trend is seen in the ALICE data on
coherent $\rho$ photoproduction in Pb-Pb UPCs at $\sqrt{s_{NN}}=2.76$ TeV~\cite{Adam:2015gsa}.
For further critical discussion of treatment of light vector meson photoproduction in ion UPCs using 
the color dipole model and Starlight Monte-Carlo generator, see Ref.~\cite{Frankfurt:2015cwa}.
%---

%---
One should note that the short Xe-Xe run did not allow one to collect high
enough statistics of events with photoproduction of quarkonia in Xe-Xe UPCs.
At the same time, studies of coherent photoproduction of vector mesons interacting with
nuclear medium with different strengths should be very informative for the precise
determination of nuclear density parameters from the transverse momentum
distribution. In particular, the position of the dips depends on the strength of absorption
of the produced mesons by the nucleus. We plan to perform in the near future such an analysis 
for nuclear targets of interest in light of future experiments at a planned electron--ion collider.

\section{Conclusion}
\label{sec:conclusion}

In this paper we presented our predictions for photoproduction of light vector mesons
in ultraperipheral Xe-Xe collisions at the LHC. We showed that the analysis of the 
data on this process
 will provide useful information on nuclear shadowing,
in particular, on the nuclear mass number $A$ dependence of nuclear shadowing in light vector meson
photoproduction on nuclei.
We argue that the measured momentum transfer distributions can be used
to gain new information on 
the density distribution of nuclear matter in $^{129}$Xe and, hence, to constrain
the elastic form factor of this nucleus, which is essential in the search for WIMP with Xenon-based
detectors.

\end{document}